# Current status of the DARPA Quantum Network

Chip Elliott[1], Alexander Colvin, David Pearson, Oleksiy Pikalo, John Schlafer, Henry Yeh
BBN Technologies, 10 Moulton Street, Cambridge MA  02138

**ABSTRACT**

This paper reports the current status of the DARPA Quantum Network, which became fully operational in BBN's laboratory in October 2003, and has been continuously running in 6 nodes operating through telecommunications fiber between Harvard University, Boston University, and BBN since June 2004. The DARPA Quantum Network is the world's first quantum cryptography network, and perhaps also the first QKD systems providing continuous operation across a metropolitan area. Four more nodes are now being added to bring the total to 10 QKD nodes. This network supports a variety of QKD technologies, including phase-modulated lasers through fiber, entanglement through fiber, and freespace QKD. We provide a basic introduction and rational for this network, discuss the February 2005 status of the various QKD hardware suites and software systems in the network, and describe our operational experience with the DARPA Quantum Network to date. We conclude with a discussion of our ongoing work.

## 1. INTRODUCTION

It now seems likely that Quantum Key Distribution (QKD) techniques can provide practical building blocks for highly secure networks, and in fact may offer valuable cryptographic services, such as unbounded secrecy lifetimes, that can be difficult to achieve by other techniques. Unfortunately, however, QKD's impressive claims for information assurance have been to date at least partly offset by a variety of limitations. For example, traditional QKD is distance limited, can only be used across a single physical channel (e.g. freespace or telecommunications fiber, but not both in series due to frequency propagation and modulation issues), and is vulnerable to disruptions such as fiber cuts or intensive eavesdropping because it relies on single points of failure.

To a surprising extent, however, these limitations can be mitigated by building QKD networks instead of the traditional, stand-alone QKD systems. Accordingly, a team from BBN Technologies, Boston University, and Harvard University has recently built and begun to operate the world's first Quantum Key Distribution network under DARPA sponsorship[2].

Detailed descriptions and rationale for this approach may be found in earlier papers[1,2,3]. This paper reports the current status of the DARPA Quantum Network, which became fully operational in BBN's laboratory in October 2003, and has been continuously running in 6 nodes operating through telecommunications fiber between Harvard University, Boston University, and BBN since June 2004. The DARPA Quantum Network is the world's first quantum cryptography network, and perhaps also the first QKD systems providing continuous operation across a metropolitan area. Four more nodes are now being added to bring the total to 10 QKD nodes.

This network supports a variety of QKD technologies, including phase-modulated lasers through fiber, entanglement through fiber, and freespace QKD. We provide a basic introduction and rational for this network, discuss the February 2005 status of the various QKD hardware suites and software systems in the network, and describe our operational experience with the DARPA Quantum Network to date. We conclude with a discussion of ongoing work.

---

[1] celliott@bbn.com, www.bbn.com, quantum.bbn.com
[2] The opinions expressed in this article are those of the authors alone, and do not necessarily reflect the views of the United States Department of Defense, DARPA, or the United States Air Force.

# 2. CURRENT QKD HARDWARE SUITES

This section describes the four different kinds of hardware suites (QKD systems) that have been integrated into the DARPA Quantum Network to date. Two have been designed and built by the BBN team for operation through telecommunications fiber: one weak-coherent system and one entanglement-based system. Two others have been built for attenuated laser pulses through freespace: one by NIST and the other by QinetiQ. We briefly discuss each in turn.

## 2.1. BBN Mark 2 Weak-Coherent System (Phase-Modulated Laser through Fiber)

BBN's Mark 2 weak-coherent system is our second-generation system that performs QKD through telecommunications fiber. As shown in Figure 1, this system uses an attenuated telecommunications laser (in the 1550 nm window) as its source, with phase-modulation provided by unbalanced Mach-Zehnder interferometers in both Alice and Bob.

The transmitter at Alice sends very dim pulses of light ("single photons") by means of very highly attenuated laser pulses at 1550.12 nm. Each pulse passes through a Mach-Zehnder interferometer at Alice and is randomly modulated to one of four phases, thus encoding both a basis and a value in that photon's self interference. The receiver at Bob contains another Mach-Zehnder interferometer, randomly set to one of two phases in order to select a basis for demodulation. The received single photons pass through Bob's interferometer to strike one of the two cooled InGaAs detectors and hence to present a received value. Alice also transmits bright pulses at 1550.92 nm, multiplexed over the same fiber, to send timing and framing information to Bob.

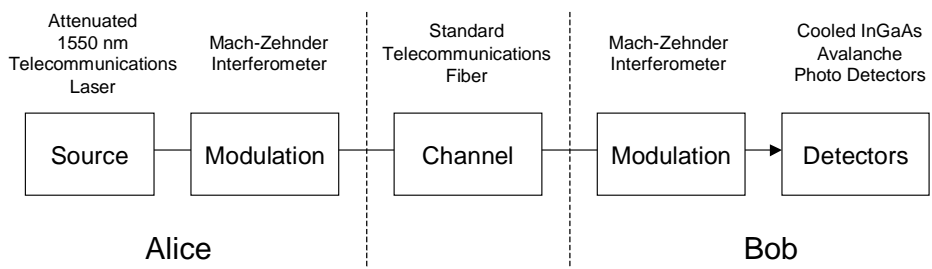

Figure 1: High-Level Schematic of BBN's Mark 2 Weak-Coherent System.

Our Mark 1 system (Alice, Bob) was implemented on optical tables but has been upgraded for compatibility with the newer Mark 2 design. The Mark 2 hardware (Anna, Boris) is rack-mounted but still employs discrete electronic components (e.g. pulse generators). If a more fully engineered system is desirable, these discrete electronics could be replaced by a single, small hardware card. The Mark 2 system is designed to operate at 5 million pulses / second. Our electronics are marginal at that rate, so we currently operate at 3.3 million pulses / second. Like many other research teams, we have selected Epitaxx EPM 239 AA Avalanche Photo Detectors (APDs) for our detectors. Detector packages include thermo-electric coolers to maintain detector operating temperatures of approximately -55 C, and incorporate hardware gating intervals and after-pulse suppression. Cooled InGaAs detector packages are supplied by IBM Almaden and by BBN, with detector Quantum Efficiencies (QE) ranging between 10% and 18% depending on the individual APDs in a given package.

The full system consists of optics, electronics that drive the optics, a realtime control computer driving the electronics, and a second computer hosting both the BBN QKD protocols and IPsec / IKE for key negotiation and traffic encryption. Figure 2 provides photographs of the four systems currently operational; please see our earlier papers for detailed equipment schematics.

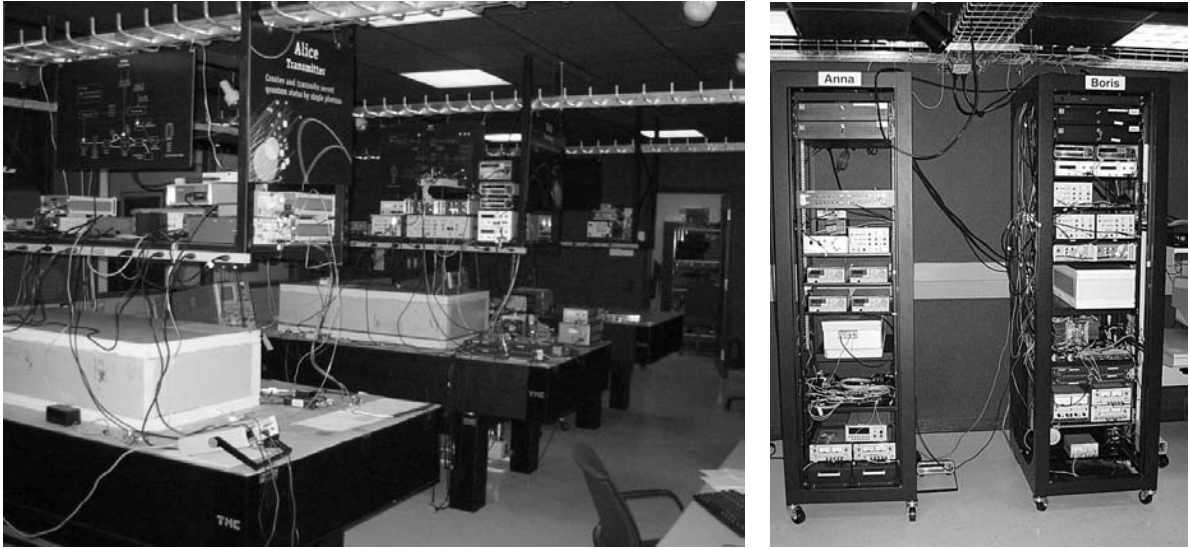

Figure 2: Equipment for BBN's Mark 2 Weak-Coherent Systems (left to right: Alice, Boris, Anna, Bob).

The Mark 2 systems can operate across a range of mean photon numbers ($\mu$). Within a room, we typically operate with $\mu = 0.1$ as is conventional for experimental systems. Across the Cambridge metropolitan area, we employ $\mu = 0.5$. Secret key yields depends on $\mu$, channel attenuation, and the exact QKD entropy estimation function employed, but these yields now range from approximately 500 bits/second between Harvard and BBN, to as many as 10,000 bits/second across the laboratory. As discussed below, the attenuation to BU is too high, and its detector quantum efficiency too low, for fully secure operation between BBN and BU. We expect to fix these issues in coming months.

**2.2. BBN/BU Mark 1 Entangled System (Polarization through Fiber)**

The BBN/BU entangled system is a BB84 system based on polarization-entangled photon pairs produced by Spontaneous Parametric Down-Conversion (SPDC). It is designed for operation through telecommunications fiber. As of February 2005, the system has been completely constructed but is not yet fully shaken down. BU's entangled source is producing coincidences and Bell-State measurements are under way. BBN's transmitter and receiver (Alex and Barb) have been built and tested with a simulated source of entangled pairs.

As this system is designed to run through telecommunications fiber, we have chosen to prepare pairs of entangled photons in the 1550 nm window for maximal distance through this fiber. Several other research teams have designed systems in which one or both nodes operate near 800 nm in order to take advantage of silicon detectors' relatively high quantum efficiency. Our design choice has a number of important consequences. One benefit is that it allows ready migration to operation in which the source of entangled photons is distant from the "transmit" and "receive" nodes. A notable drawback, however, is the relatively low Quantum Efficiency (QE) of InGaAs APDs. Since entanglement-based cryptography requires the correlated detection of two photons, the raw (unsifted) bit rate is limited by the square of a single APD's detection probability.

The Mark 1 Entangled system employs polarization modulation to encode the (basis, value) pairs needed for quantum cryptography. Such modulations can be produced relatively easily in Alex and Barb. In fact, the random selection of the value happens in the pair generation process, and random selection of basis can be performed purely passively by interposition of a beam splitter. This simplicity stands in contrast with the relative complexity of phase modulation, which requires carefully tuned Mach-Zehnder interferometers and an external source of randomness that drives deterministic phase modulators. However, polarization is quite difficult to transmit through a telecommunications fiber, which generally acts as a "polarization scrambler." Thus an important part of the Mark 1 Entangled System design is its polarization control.

At present, the Mark 1 Entangled System runs BB84 cryptographic protocols, rather than the Ekert protocols. We do this because we already have a well-debugged version of BBN's BB84 protocol stack, which we can directly employ on this new system. In later stages of this project, we may also implement the Ekert protocols. Figure 3 provides a schematic of this system. As shown, the source is currently "within" Alex. Alex receives a randomly polarized photon from each entangled pair, and transmits its twin to Barb. Alex's photon is further processed so that it will strike one of Alex's 4 QKD Cooled APDs thus encoding a random (basis, value) selection at Alex. This detected result is then passed to Alex's electronics as input to the BB84 protocols.

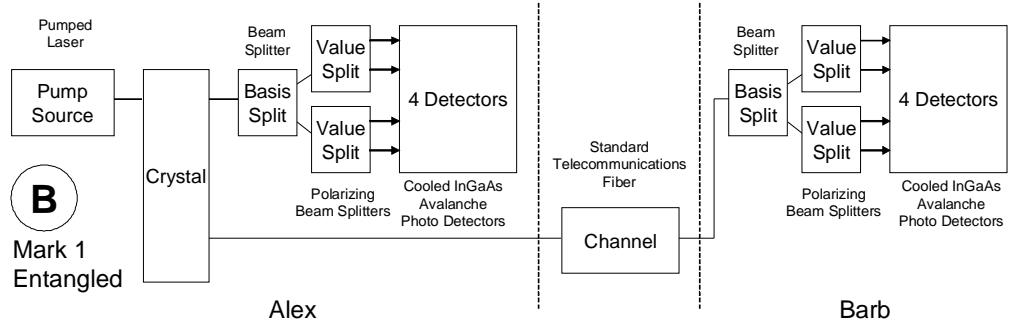

Figure 3: High-Level Schematic of the Mark 1 Entangled System.

On the receiver side, Barb contains a detector suite identical to Alex's. This includes four 1550 nm QKD detectors which are operated in the Geiger gated mode, where the applied bias voltage exceeds the breakdown voltage for a very short period of time when a photon is expected to arrive, leading an absorbed photon to trigger an electron avalanche. Since the typical gating interval is ~ 1 ns, this mode of operation requires some knowledge of the photon arrival time, which is deduced from an incoming synchronization pulse from Alex (not shown).

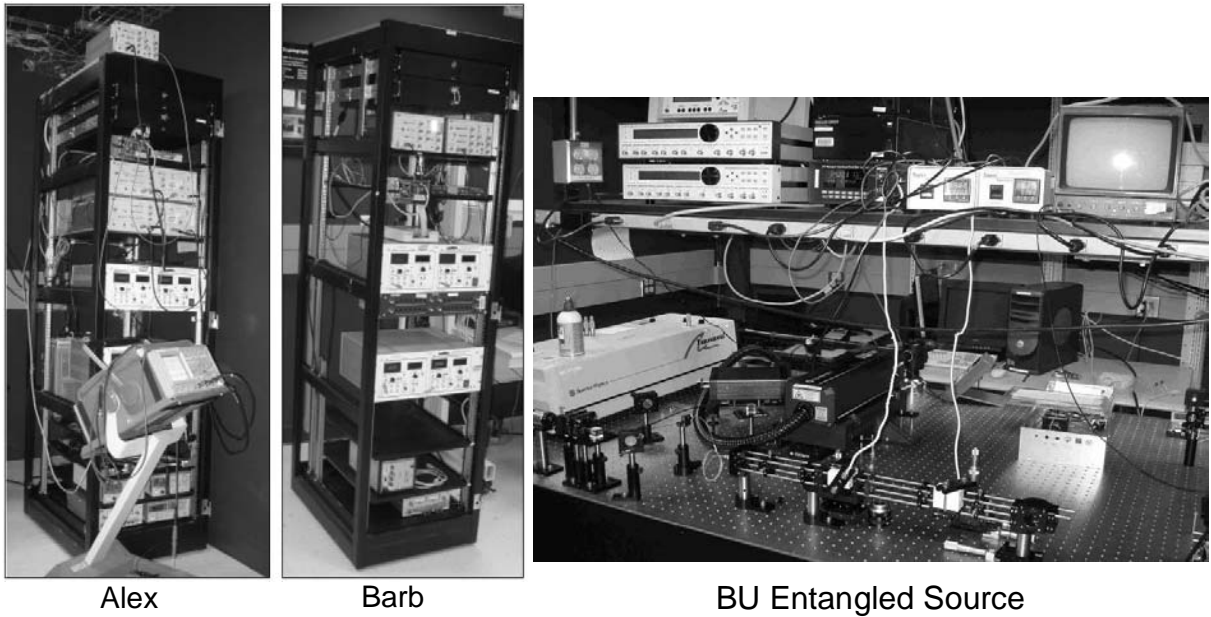

Figure 4: Equipment for the Mark 1 Entangled System.

Figure 4 provides photographs of the Alex and Barb electronics and optics, and BU's source of entangled photons. As can be seen, the BU source is currently a collinear arrangement of two Periodically Poled Lithium Niobate (PPLN) crystals driven by a pulsed pump source.

In idealized operation, Alex and Barb receive identical (basis, value) signals in their electronics for each entangled pulse. Practice is of course different from idealized theory. First, detector efficiencies are relatively low (e.g. 15 – 20%), and the series of receive beamsplitters and filters causes attenuation. Thus often Alex will not detect her photon. Barb will have an even worse time, of course, since there is considerable attenuation through the fiber channel before Barb's receiver suite. Note that both Alex and Barb must detect a given pulse before BB84 can even begin to operate. Second, detector dark count will induce spurious detection events, and thus give rise to occasional spurious correlations between detections at Alex and Barb, thus raising the Quantum Bit Error Rate (QBER) of the system as a whole.

### 2.3. NIST Freespace System (Polarization)

The third system present in the DARPA Quantum Network is a very high-speed freespace system designed and built by a team from the National Institute of Standards, led by Dr. Carl Williams and Dr. Joshua Bienfang. At present only the electronics for this system are set up in BBN's Cambridge laboratory, and the two nodes (Ali and Baba) are connected by direct coaxial cabling instead of the true freespace quantum and public channels used between the corresponding nodes at NIST. BBN's software has been fully integrated onto these electronics, however, and has been operational for nearly a year. In collaboration with the NIST team, we expect to build exact copies of the full NIST system, e.g., its optics and telescopes, during the spring of 2005 and to start running through the atmosphere across BBN's campus.

Figure 5 shows a photograph of the Ali and Bab electronics for NIST's quantum channel, as installed in BBN's laboratory. The two devices are connected by coaxial cabling, and monitored by an attached display. BBN's QKD software resides in both equipment chasses, along with NIST firmware and hardware.

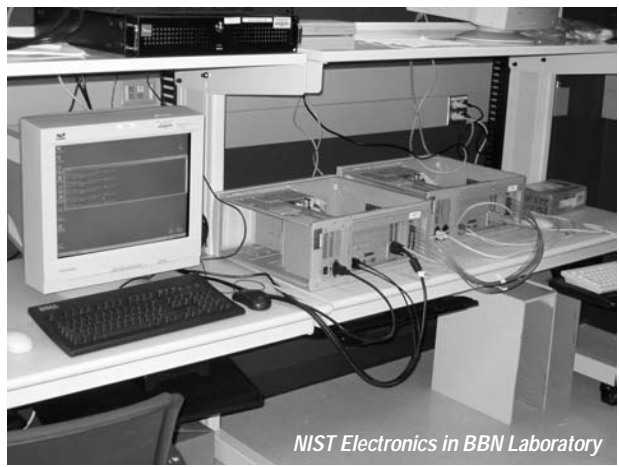

Figure 5: Electronics for NIST's Freespace System in BBN Laboratory.

The NIST team has provided detailed technical explanations of their system[4]. Salient points are repeated here for ease of reference. The quantum channel at 845 nm operates in parallel with a classical channel at 1550 nm. The system is designed to operate at 1.25 billion pulses / second, making it far faster than other existing systems. At present timing jitter in the silicon detectors for the quantum channel limit operation to perhaps one third of this design rate, but these issues will likely be resolved in the near future. This system has been demonstrated in operation over a 730 meter atmospheric channel between two NIST buildings.

NIST's four quantum sources are 10 GHz vertical-cavity surface-emitting lasers (VCSELs) with bias voltage set to produce 250 ps pulses with high extinction ratio. These pulses are attenuated with variable fiber attenuators and then coupled, via single-mode fiber, to free-space optics mounted on the back of the transmit telescope where they are collimated, linearly polarized in either the vertical or +45 degree direction, and then combined with a non-polarizing beam-splitting cube. The beam is then shaped to fill the entire 20.3 cm diameter output aperture of the Schmidt-Cassegrain telescope. The receive telescope is identical to the transmit telescope. A non-polarizing beam-splitting cube at the output of the receive telescope performs Bob's random choice of polarization basis. The 'value' measurements are made with polarizing beam-splitting cubes, and then fiber-coupled via to a detector box, where they pass through a 2 nm spectral filter, and finally focused on to silicon avalanche photodiodes (APDs). Polarization extinction ratios in excess of 500:1 are observed for both bases.

### 2.4. QinetiQ Freespace System (Polarization)

The QinetiQ freespace system is a small, portable QKD system designed for easy operation through the atmosphere. Since it will be described in detail by its designers in a companion paper[5] in this SPIE workshop, we do not outline any background technical information here. In October 2004, the BBN and QinetiQ teams integrated BBN's QKD software onto the QinetiQ hardware suite at QinetiQ for a one-off compatibility test. All appeared to work well, and we expect delivery of QinetiQ hardware to BBN's Cambridge laboratory in the spring of 2005 for final integration and full incorporation into the DARPA Quantum Network.

## 3. CURRENT STATUS OF BBN's QKD SOFTWARE

This section describes the BB84 software currently implemented in the DARPA Quantum Network. (We omit discussion of important ancillary software such as key management and usage, network management, etc., due to space limitations. We also omit discussions of BBN's cryptographic key routing and key relay protocols as beyond the scope of this paper.) As will be seen, our general approach has been to implement every major algorithm for QKD rather than to try to pick the "best" approach. Furthermore, the same software runs on all the QKD Hardware Suites described above. A simple configuration setting indicates which selection of algorithms will be used during a given period of operation on a specific QKD system.

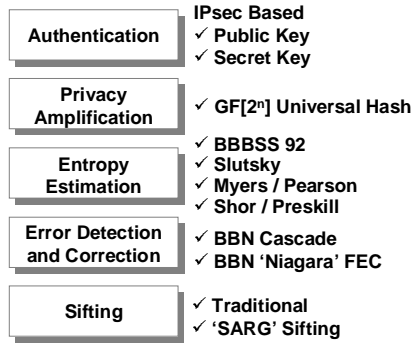

Figure 6: Subsystems in BBN's QKD Protocols with Current Implementations.

Figure 6 presents in schematic form the basic steps required for QKD, reading upwards from the bottom as is typical of networking protocol stacks, along with the current techniques implemented for each step. We discuss each of these processes in further detail below.

### 3.1. Sifting

Sifting allows Alice and Bob reconcile their "raw" secret bit streams to remove the following types of errors:

- No detection symbols, where Alice believed that she prepared and transmitted a photon but none of Bob's detectors fired. Such events are very common and have several distinct causes. Some of the most common are: (a) Alice's source did not actually emit a photon; (b) that photon was lost in transmission; (c) Eve captured the photon and did not replace it; (d) Bob's detector didn't fire when the photon hit it.

- Wrong basis symbols, where Bob did receive a detection but in fact Alice and Bob did not use the same basis for this bit, i.e., the basis that Alice randomly selected was different from that which Bob randomly selected.

- Multiple detection symbols, in which more than one of Bob's detectors fired. This makes it impossible for Bob to determine whether the symbol should be treated as a 0 or a 1, and so the symbol must be discarded.

The sifting process involves both communication (QKD public protocol interactions) between Alice and Bob, and internal algorithms within Alice and Bob. After sifting, Alice and Bob discard all other symbols and keep only the resulting "sifted" bits.

The DARPA Quantum Network implements two forms of sifting: "classic" and SARG04[6]. In classic sifting, either Alice or Bob may announce the basis choice openly, and in fact Eve may be permitted to overhear it. In SARG sifting, the basis choice itself is not revealed. Instead, two possibilities are revealed, only one of which is compatible with the idealized detection event, resulting in less basis knowledge for Eve.

### 3.2. Error Detection and Correction

Once Alice and Bob have agreed on sifted bits, they must perform error correction to find and then eliminate those bits that have been damaged in transmission, i.e., the flipped bits sent as a 1 but received as a 0 or vice versa. There are many ways to perform error correction but each has these important consequences:

1. Error correction is always probabilistic – unless all bits are revealed during the process! -- and thus there is some possibility that Alice and Bob will believe that they share an identical set of bits, but in fact they do not.

2. Error correction requires communications between Alice and Bob, and inevitably – assuming that Eve can obtain plain text versions of all such public communications – the process of error correction reveals information about the sifted bits to Eve.

3. The process of error detection allows Alice and Bob to estimate the current Quantum Bit Error Rate (QBER) on the quantum channel between them, which can then be used as input for privacy amplification.

Thus the end result of error correction is three-fold. First, Alice and Bob will with high probability contain, in their local memories, identical copies of a set of error-corrected bits. Second, Eve will have some knowledge of these bits' values, which must then be reduced by the subsequent process of privacy amplification. Third, Alice and Bob may estimate the current QBER from the results of this process.

The DARPA Quantum Network currently implements two forms of error detection and correction: a modified version of the well-known Cascade protocol, and a new Forward Error Correction (FEC) technique that we term "Niagara."

**BBN Cascade**. Brassard and Salvail's Cascade protocol[7] is the first and best-known error correction protocol for QKD. It usually performs within about 15% to 20% of the optimal number of bits revealed (Shannon limit), and while it does require an initial estimate of the error rate, it can adapt fairly well if the error rate deviates from this. It is also adapts gracefully to different sizes of input message. In our view, Cascade is efficient enough for relatively low-rate QKD systems, dealing with sifted key rates of less than 50,000 bits per second. BBN Cascade is an implementation that has been optimized for protocol efficiency.

**BBN Niagara**[8]. This BBN-designed protocol is a novel type of Low-Density Parity Check (LDPC) code designed for use in QKD applications. It is a form of Forward Error Correction, i.e., it does not require multiple protocol interactions between Alice and Bob as does Cascade.

Like most forward error-correcting codes, Niagara is based on parity checks. In such schemes, the initial message M (a bit vector) is multiplied by a generator matrix G over GF[2] to yield a longer encoded message C. At the receiver, C is multiplied by a parity-check matrix H over GF[2], and if the result is the zero vector, the message is accepted as correct. An LDPC code is one in which the check matrix H is sparse. Sparseness does not make the code any more capable of detecting or correcting errors, but does make decoding more tractable. The iterative decoding algorithm used, the sum-product algorithm, requires time proportional to the number of one bits in the parity check matrix times the number of iterations needed, which is generally fixed at some maximum, e.g. 20, after which the algorithm declares failure.

Proper code construction is essential for good performance from LDPC codes. Niagara's input parameters are the block size *b* (the number of data bits), the number of parity bits *p* to be revealed, the density of the one bits in the parity-check matrix, and a seed for a pseudo-random number generator. These parameters uniquely specify a code matrix, allowing both parties to construct identical codes with minimal communication over the public channel. At present we fix the density so that each data bit is involved in an average of 5 parity checks; as a result, the density need not be sent to the other party. To choose the number of parity bits *p*, we take the recent error rate *e* on the channel, add a noise margin for variation in the error rate, convert this to a code rate using Shannon's channel-coding theorem, then add another margin for sub-optimal codes. (See Figure 7.) This gives a total number of parity bits about 25% above the Shannon limit, with very little decoding failure.

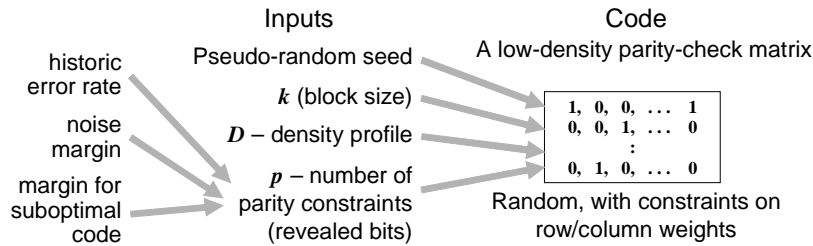

Figure 7: Inputs to BBN's "Niagara" Low-Density Parity Check (LDPC) Codes for QKD.

Figure 8 provides an example of the performance improvements achieved by the Niagara protocol over BBN's version of the Cascade protocol. As can be seen, it greatly reduces the communications overhead and delay for error correction, and also greatly reduces the CPU usage required, while imposing only a modest expense in coding efficiency as compared to the Shannon limit.

| Average for 4096 bit blocks, 3% error rate | BBN Cascade | BBN Niagara |
|---|---|---|
| Revealed bits | 958 | 1006 |
| % of Shannon limit | 120% | 126% |
| Delay (round trips) | 68 | 1 |
| Communication (bytes) | 19200 | **480** |
| CPU usage (secs / Mb, 800MHz x86) | 17.4 | 1.1 |

Figure 8: Example Performance Comparison of BBN Cascade and BBN Niagara Implementations.

### 3.3. Entropy Estimation

Privacy amplification depends on having an accurate estimate of the eavesdropping-free entropy sifted and error corrected secret bit sequences—i.e., the amount of entropy in these bits beyond what Eve might know. This estimation process is not usually broken out as a separate step, but since it is crucial for the security of quantum cryptography, we treat it as a distinct operation in QKD processing.

At present, the QKD community does not seem to have a universally agreed-upon calculation for this entropy. Hence the DARPA Quantum Network implements four different entropy estimation techniques, which we term Bennett, Slutsky, Myers-Pearson, and Shor-Preskill. A full treatment of these techniques and their underlying assumptions is well beyond the scope of this paper but Figure 9 presents the formulae used for each.

How much usable information is in the corrected QKD bits beyond what Eve may know, given:

- **b**, the number of received bits (sifted)
- **e**, the number of errors in the sifted bits
- **n**, the total number of bits transmitted
- **d**, the number of parity bits disclosed during error correction
- **r**, a non-randomness measure from randomness tests
- **c**, a confidence parameter (chance of underestimating)

The DARPA Quantum Network implements four measures:

**Bennett, et al.:** $t = b - e - \frac{4e}{\sqrt{2}} - \text{erf}^{-1}(1-c)\sqrt{(8+4\sqrt{2})e}$

**Slutsky, et al.:** $t = (b-e)\left[1 + \log_2\left(1 - \frac{1}{2}\left(\frac{1-3e'}{1-e'}\right)^2\right)\right] - \text{erf}^{-1}(1-c)\sqrt{\frac{b-e}{2}}$ where $e' = \min\left(\frac{e}{b} + \frac{\text{erf}^{-1}(1-c)}{\sqrt{2b}}, \frac{1}{3}\right)$

**Myers / Pearson:** $t = \max_{R \in (1,2]}\left[(b-e)\frac{b-e}{1-R}\log_2\left(p_E^R + (1-p_E)^R\right) - \log_2\left(\frac{R}{c(R-1)}\right) - 2\right]$ where

$p_E = \frac{1}{2} + \sqrt{\frac{p}{1-p}\left(1 - \frac{p}{1-p}\right)}$ and $p$ solves $\sum_{i=0}^{e}\binom{b}{i}p^i(1-p)^{n-i} = c$

**Shor / Preskill:** $t = (b-e)(1 + p\log_2(p) + (1-p)\log_2(1-p)) + 2\log_2(c)$, $p$ as above

**For all, the entropy estimate is:** $t - r - d - m_1 n - m_2 b$

Figure 9: Currently Implemented QKD Entropy Estimation Functions in the DARPA Quantum Network.

We note that the choice of entropy estimation function has extremely important practical consequences. Of greatest importance is the fact that an incorrect estimate may lead to insufficient privacy amplification, and thus allow Eve to know more about the resultant "secret" bits than expected. Even when providing the same estimate asymptotically, the different techniques also have different efficiency implications for the finite-sized blocks of bits used in practical QKD systems. Figure 10 displays these efficiencies for the four entropy estimation techniques we currently implement.

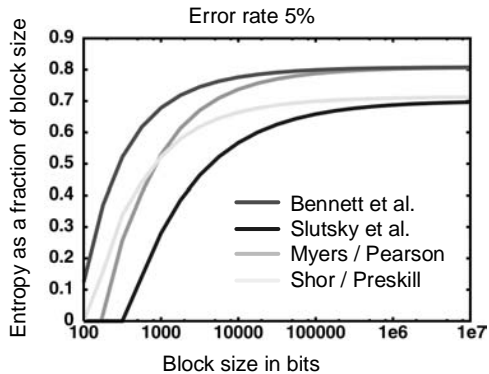

Figure 10: Example Performance Comparison of Entropy Estimation Functions on Finite Block Sizes.

### 3.4. Privacy Amplification

Privacy amplification is the process whereby Alice and Bob reduce Eve's knowledge of their shared bits to an acceptable level. It is also often called distillation or advantage distillation. Please recall that Eve may acquire her illicit knowledge in several ways, such as by eavesdropping and/or by observing bits exchanged in the process of error correction. In some cases, Eve may actually know the exact values of specific bits held by Alice and Bob; but more likely, Eve has only some indications that certain families of sequences are more probable than others.

Privacy amplification is a purely classical algorithm that operates on bits in computer memory, and that "smears out" the value of each initial shared bit across the shorter resulting set of bits. Assuming that the "smearing" works properly, one can reduce Eve's knowledge as desired by picking the size of the resulting set of bits. The shorter the resulting set of bits, the less Eve knows. Unfortunately, though, Eve's knowledge can never be reduced to zero by this technique – unless one throws away all the bits! Still, Eve's knowledge can be made arbitrarily small.

In the DARPA Quantum Network, the QKD node initiating privacy amplification selects a linear hash function over the Galois Field $GF[2^n]$ where $n$ is the number of error-corrected bits in a block. It then transmits four items to the other end—the number of bits $m$ of the shortened result, the (sparse) primitive polynomial of the Galois field, a multiplier ($n$ bits long), and an $m$-bit polynomial to add (i.e. a bit string to exclusive-or) with the product. Each side then performs the corresponding hash and truncates the result to $m$ bits to perform privacy amplification.

### 3.5. Authentication

Authentication is the process whereby Alice and Bob assure themselves that, with very high probability, they are really exchanging information with each other and not with Eve. Such authentication must be *mutual* so that Alice has high assurance that she is indeed talking to Bob, and Bob has similar assurance about Alice. It must also be *continuous* so that all aspects of the ongoing interactions between Alice and Bob are protected, and not just an initial handshake.

Our approach to authentication follows the original prescription for authentication in quantum cryptography, namely, Universal Hash Functions. This basic approach was very clearly described in the BB84 paper:

> "The need for the public (non-quantum) channel in this scheme to be immune to active eavesdropping can be relaxed if the Alice and Bob have agreed beforehand on a small secret key, which they use to create Wegman-Carter authentication tags for their messages over the public channel. In more detail the Wegman-Carter multiple-message authentication scheme uses a small random key to produce a message-dependent 'tag' (rather like a check sum) for an arbitrary large message, in such a way that an eavesdropper ignorant of the key has only a small probability of being able to generate any other valid message-tag pairs. The tag thus provides evidence that the message is legitimate, and was not generated or altered by someone ignorant of the key. (Key bits are gradually used up in the Wegman-Carter scheme, and cannot be reused without compromising the system's provable security; however, in the present application, these key bits can be replaced by fresh random bits successfully transmitted through the quantum channel.)"

In accordance with our general philosophy that QKD forms a *part* of an overall cryptographic architecture, and not an entirely novel architecture of its own, the DARPA Quantum Network currently employs the standardized authentication mechanisms built into the Internet security architecture (IPsec), and in particular those provided by the Internet Key Exchange (IKE) protocol. These mechanisms currently allow for all control message traffic (QKD protocols) to be authenticated by public keys in the nodes, and/or by pre-placed secret keys. Our plan is to extend this architecture by further incorporating those BB84 Universal Hash Functions described above in order to achieve continuous authentication based on secret bits derived from ongoing QKD.

## 4. CURRENT STATUS OF THE CAMBRIDGE-AREA NETWORK

Figure 11 displays a fiber map of the DARPA Quantum Network's buildout through Cambridge, Mass., as of February 2005. The network consists of two weak-coherent BB84 transmitters (Alice, Anna), two compatible receivers (Bob, Boris), and a 2x2 switch that can couple any transmitter to any receiver under program control. Alice, Bob, and the switch are in BBN's laboratory; Anna is at Harvard; and Boris is at Boston University. The fiber strands linking Alice, Bob, and the switch are several meters long. The Harvard-BBN strand is approximately 10 km. The BU-BBN strand is approximately 19 km. Thus the Harvard-BU path, through the switch at BBN, is approximately 29 km. All strands are standard SMF-28 telecommunications fiber.

Figure 12 presents the distances and attenuations for the current fiber spans across Cambridge. As can be seen, the BBN-Harvard spans are 10.2 km (5.1 dB) and the BBN-BU spans are 19.6 km (11.5 dB). These attenuations are quite high, being equivalent to 24.3 km and 54.8 km respectively of standard fiber at 0.21 dB/km, and are incurred by a large number of connectors along the current fiber paths. A sample Optical Time Domain Reflectometry (OTDR) trace for the BU-BBN path, at the left, illustrates the effects of these connectors.

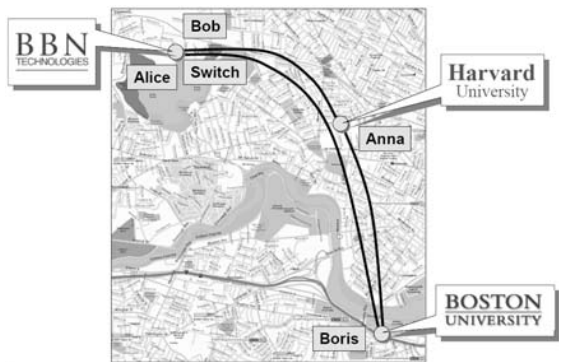 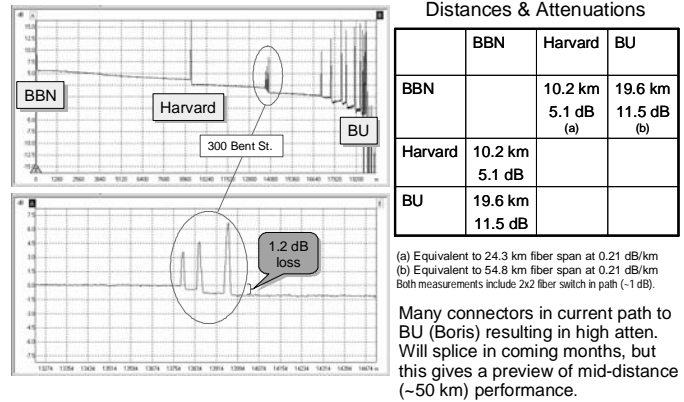

Figure 11: Logical Map of the Cambridge-Area Fiber Network.

Figure 12: Practical Considerations of Attenuation in Cambridge-Area Network due to Connectors.

Anna's mean photon number is 0.5 at present, with the Anna-Bob system delivering about 1,000 privacy-amplified secret bits/second at an average 3% Quantum Bit Error Rate (QBER). Right now the DARPA Quantum Network cannot support fractional mean photon numbers to Boris at BU, due to high attenuation in fiber segments across the Boston University campus and relatively inefficient detectors in Boris. (BBN-BU attenuation is approximately 11.5 dB). Thus the network currently operates at a mean photon number of 1.0 on the BBN-BU link, in order to continuously exercise all parts of the system, even though the resultant secret key yield is zero. In the near future, fiber splices and perhaps detector upgrades should allow operation to BU with mean photon numbers of 0.5.

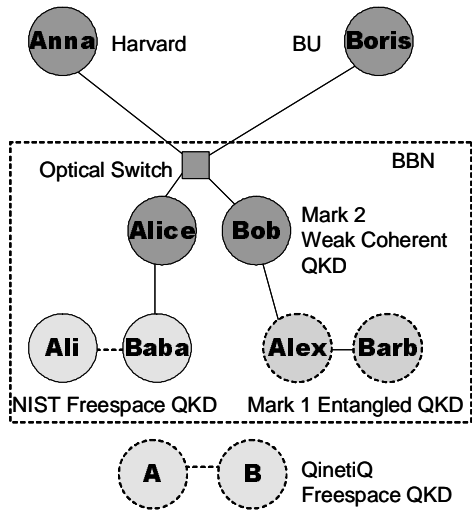

Figure 13: Current Topology of the DARPA Quantum Network.

As shown in Figure 13, in addition to these fiber-based systems, the DARPA Quantum Network also contains Ali and Baba, the electronics subsystems for a high-speed freespace QKD system designed and built by the National Institute of Standards and Technology (NIST). Ali and Baba run the BBN QKD protocols, and are linked into the overall network by BBN's QKD key relay protocols between Ali and Alice. It further contains two new entanglement-based nodes named Alex and Barb, built jointly by BU and BBN, but these nodes are not yet fully operational. In the near future, it will also link in two new freespace nodes provided by QinetiQ, which have yet to be named.

# 5. ONGOING WORK

Near-term work on the DARPA Quantum Network falls into three main categories:

- Shakedown of Entangled System. Once the BU source has been adequately characterized, we will shake down the full Entangled System and begin its routine operation in the network.

- Full integration of NIST and QinetiQ Freespace Systems. Over the coming months, we expect to acquire all necessary optics for these two systems, shake them down in the BBN laboratory, and then begin continuous operation across the BBN campus.

- Development of a Superconducting Single-Photon Detector (SSPD). Rochester University, NIST Boulder, and BBN are collaborating on building an ultra-fast single-photon detector for telecom wavelengths. We plan to have the first proof-of-concept system operating within the coming year.

# 6. ACKNOWLEDGEMENTS AND REFERENCES

We are deeply indebted to Dr. Jonathan Smith (DARPA IPTO) and Dr. Don Nicholson (Air Force Research Laboratory) who are the sponsor and agent, respectively, for this research project, and to Dr. Mike Foster, the original DARPA sponsor of this work. This paper reflects highly collaborative work between the project members. Of these, particular credit is due to Profs. Alexander Sergienko and Gregg Jaeger, and Dr. Martin Jaspan (Boston University), Dr. John Myers and Prof. Tai Wu (Harvard), and Alex Colvin, John Lowry, David Pearson, Oleksiy Pikalo, John Schlafer, Greg Troxel, and Henry Yeh (BBN).

As anyone who works will fiber-based QKD can attest, detectors are the most critical technology at present. Although BBN built its own set of cooled InGaAs APDs for our first two years of development, we have been very fortunate to switch to superb InGaAs detector systems built by Dr. Don Bethune and Dr. William Risk at IBM Almaden for our fiber-based systems. We are extremely grateful for such fine engineering on their part. We also thank Dr. Carl Williams and Dr. Joshua Bienfang and their team at NIST for their generous, long-term loan of their QKD system, and to P. R. Tapster, P. M. Gorman, D. M. Benton, D. M. Taylor, B. S. Lowans of Qinetiq for their development of a freespace QKD system for our network. Our interest in QKD networks was sparked by the prior work of, and discussions with, the quantum cryptography groups at IBM Almaden and Los Alamos, and by the kind hospitality of Dr. David Murley several years ago.